# Air Stable and Layer Dependent Ferromagnetism in Atomically Thin van der Waals CrPS$_4$

Joolee Son[1,2†], Suhan Son[1,3,4†], Pyeongjae Park[1,3,4], Maengsuk Kim[5], Zui Tao[6], Juhyun Oh[7], Taehyeon Lee[1], Sanghyun Lee[8], Junghyun Kim[1,3,4], Kaixuan Zhang[1,3,4], Kwanghee Cho[9], Takashi Kamiyama[9], Jun Hee Lee[5], Kin Fai Mak[6], Jie Shan[6], Miyoung Kim[7], Je-Geun Park[1,3,4*], and Jieun Lee[1,2*]

[1]Institute of Applied Physics and Department of Physics and Astronomy, Seoul National University, Seoul 08826, Korea, [2]Department of Physics and Department of Energy Systems Research, Ajou University, Suwon 16499, Korea, [3]Center for Quantum Materials, Seoul National University, Seoul 08826, Korea, [4]Center for Correlated Electron Systems, Institute for Basic Science, Seoul 08826, Korea, [5]School of Energy and Chemical Engineering, Ulsan National Institute of Science and Technology, Ulsan 44919, Korea, [6]Department of Physics and School of Applied and Engineering Physics, Cornell University, Ithaca, NY 14850, USA, [7]Department of Materials Science and Engineering and Research Institute of Advanced Materials, Seoul National University, Seoul 08826, Korea, [8]LG Chem Research Park, Daejeon 34122, Korea, [9]Materials and Life Science Division, J-PARC Center, Japan Atomic Energy Agency, Tokai 319-1195, Japan

† These authors contributed equally to this work.

* Correspondence to: jgpark10@snu.ac.kr (J.-G.P.) and lee.jieun@snu.ac.kr (J.L.)

**ABSTRACT:** Ferromagnetism in two-dimensional materials presents a promising platform for the development of ultrathin spintronic devices with advanced functionalities. Recently discovered ferromagnetic van der Waals crystals such as $CrI_3$, readily isolated two-dimensional crystals, are highly tunable through external fields or structural modifications. However, there remains a challenge because of material instability under air exposure. Here, we report the observation of an air stable and layer dependent ferromagnetic (FM) van der Waals crystal, $CrPS_4$, using magneto-optic Kerr effect microscopy. In contrast to the antiferromagnetic (AFM) bulk, the FM out-of-plane spin orientation is found in the monolayer crystal. Furthermore, alternating AFM and FM properties observed in even and odd layers suggest robust antiferromagnetic exchange interactions between layers. The observed ferromagnetism in these crystals remains resilient even after the air exposure of about a day, providing possibilities for the practical applications of van der Waals spintronics.

Since the recent introduction of two-dimensional (2D) van der Waals ferromagnets, the feasibility of implementing 2D crystals for the development of spin-dependent electronics, optoelectronics, and quantum information devices has rapidly increased.[1-16] In particular, layer-dependent ferromagnetism recently discovered in 2D materials such as $CrI_3$ offers fascinating opportunities for the development of spintronic device applications because of wide tunabilities using doping, electric field, light, and pressure.[8-12,15] With these tunabilities and the ability to assemble various 2D materials in vertical heterostructures, van der Waals spin logic devices have been demonstrated including spin filters based on giant tunneling magnetoresistance.[14] However, the scope of the device geometry and functional characterization have been limited because of the material's degradation in air through hydration or oxidation.[7,13] Although inert gas environment using glove box and capping layer such as hexagonal boron nitride and graphene have been employed to delay the degradation, it is crucial to identify fundamentally air stable 2D ferromagnetic materials to explore further possibilities.

$CrPS_4$ is a candidate magnetic van der Waals material which can be exfoliated down to a monolayer limit with intrinsic spin orderings. Because of the material's multi-bonded crystal structure and chemical composition, the material could potentially be air stable. The crystal in the bulk form is found to be an antiferromagnet,[17] but there has been a controversial on the magnetic ground state of the monolayer.[18,19] Therefore, it is critical to resolve the local magnetic alignments in the monolayer and few-layer $CrPS_4$. Furthermore, the crystal structure of $CrPS_4$ constituted by chains of chromium octahedra ($CrS_6$) interconnected by phosphorous in the 2D rectangular lattice[20] (Figure 1a) suggests the intrinsic in-plane anisotropy which can allow a wide variety of magnetic ground states in the crystal.[19] Revealing the magnetic ground state in thin layer $CrPS_4$, therefore, provides an interesting testbed to study rich magnetic anisotropy in 2D materials, adding interesting opportunities for the development of 2D spin-dependent device applications.

In this work, we report the local spin ground states of monolayer, bilayer, and few-layer $CrPS_4$. $CrPS_4$ in the bulk form is found to be an A-type antiferromagnet as revealed by the neutron powder diffraction experiment. When thinned down to a monolayer, a ferromagnetic ground state is detected with an out-of-plane spin orientation using the magneto-optic Kerr effect microscopy. In the intermediate thicknesses, even layers are found to be antiferromagnetic (AFM), but odd layers are ferromagnetic (FM). The layer-dependent alternation of FM-AFM spin ground states agrees well with the bulk form, suggesting that exchange interactions in thin CrPS$_4$ are retained after exfoliation down to the monolayer. Moreover, the FM ground states observed in odd layers are found to remain intact after about a day of air exposure, demonstrating excellent material stability advantageous for investigating other heterostructures and devices.

## RESULTS AND DISCUSSION

The bulk crystal in our work was prepared by using the chemical vapour transfer technique (details given in Method). The magnetic susceptibilities measured under the magnetic field of 100 Oe on the bulk CrPS$_4$ along and perpendicular to the c-axis show that the material is an antiferromagnet with an out-of-plane spin orientation (Figure 1b). However, by taking the inverse susceptibility of CrPS$_4$ perpendicular to the c-axis ($\chi_\perp$), a significant positive Curie-Weiss temperature of +41 K is observed (Figure 1c). It strongly implies that the strongest exchange interaction in this material is ferromagnetic. On the other hand, most 2D materials have relatively weak van der Waals couplings, leading to the small exchange interactions through the interlayer. These observations suggest that CrPS$_4$ is likely to be ferromagnetically coupled within the layer with an interlayer antiferromagnetic order, *i.e.,* an A-type AFM ordering.

To confirm the A-type antiferromagnetic ground state directly, we carried out the time-of-flight neutron powder diffraction. Figure 1d depicts the measured neutron diffraction pattern. By using the C2/m space group of CrPS$_4$, we tested four different symmetry allowed magnetic configurations in the system and located the most probable spin configuration by Rietveld analysis. We unambiguously found that our experimental data are consistent with an A-type AFM ground state with an ordered magnetic moment of about 3 $\mu_B$/Cr$^{3+}$, in agreement with recent reports.[21,22]

More details can be found in Supporting Information Fig. S1 and S2. The magnetic ground state of bulk in the absence of an external field is illustrated in the inset of Figure 1d.

Next, for the investigation of the magnetic properties of monolayer, bilayer, and few-layer $CrPS_4$, we obtained thin $CrPS_4$ by exfoliating bulk crystals onto oxidized silicon substrates. The exfoliation is performed in air unless otherwise noted. Figure 2a shows the microscope image of 1L to 5L flakes that are obtained on a single substrate. The layer number of each flake is determined by atomic force microscopy using the monolayer thickness of about ~ 6 Å (Supporting Information Fig. S3). The layer number is also checked by site-dependent Raman spectra shown in Figure 2b. The Raman peaks are labeled alphabetically from A to L in the range between 100 and 400 cm$^{-1}$. Note that the intensity of the F peak increases linearly as a function of the layer number (Figure 2d). Moreover, the frequency of the B peak is found to monotonically redshift with the increasing thickness, which agrees well with literature[23] and further confirms the layer thickness (Supporting Information Fig. S7).

With the information on the layer number, the magnetic properties of thin $CrPS_4$ are probed by spatially resolved magneto-optic Kerr effect (MOKE) microscopy. In our experimental set-up, a linearly polarized probe light propagating in the normal direction to the sample plane is employed, which allows the detection of the out-of-plane spin magnetization.[24,25] Generally, the MOKE in this geometry probes the out-of-plane spin orientation given that the material is optically isotropic. $CrPS_4$, on the other hand, is a birefringent material having the monoclinic crystal structure with two optical axes.[23] Such optical anisotropy can complicate the separate measurement of the MOKE signal due to the birefringent optical rotation. To remove the contribution from the birefringence, we measured the optical rotation as a function of the probe polarization angle relative to the crystal orientation as shown in Figure 1e at room temperature and identified two optical axes (a- and b-axis) (Supporting Information Fig. S8). For magnetization measurement, we chose the probe beam polarization to be parallel to one of the a- or b-axis, where the optical rotation by birefringence is minimal. With the aligned probe polarization, the MOKE measured on a ~200 nm thick CrPS$_4$ exhibits negligible Kerr rotation angle ($\theta_{KR}$) in the temperature range between 10 and 53 K, confirming zero net magnetization of bulk (Figure 1f).

The spin orientation of thin layer CrPS$_4$ is then measured for the sample shown in Figure 2a. The constant magnetic field of 0.25 T along the out-of-plane direction is applied throughout the measurement. As shown in Figure 2c, in 2L and 4L flakes, null $\theta_{KR}$ is detected in the given temperature range, similar to bulk. Interestingly, in 1L, 3L, and 5L flakes, nonzero $\theta_{KR}$ are detected at temperatures below 23 K, which is a hallmark of the out-of-plane spin orientation. The same temperature dependence is also observed using magnetic circular dichroism (MCD), further confirming the existence of the net out-of-plane spin magnetization in odd layers (Supporting Information Fig. S10). To obtain the transition temperature ($T_C$), the experimental data is fitted with the power-law using $\theta_{KR}(T) = \alpha(1 - T/T_C)^{\beta}$, where $\alpha$ and $\beta$ are proportional constant and critical exponent, respectively.[25,26] The fit parameters summarized in Figure 2e show that the critical exponents of 5L and 3L flakes are found to be ~ 0.4 while in 1L the value reduces to ~ 0.2, suggesting that CrPS$_4$ in the monolayer form is close to a 2D Ising model system. On the other hand, the measured $T_C$ showed little dependence on layer thickness, implying relatively weak spin interaction in vertical direction compared to in-plane interactions in CrPS$_4$.

Similar layer dependence of magnetization is consistently observed in the 2D scanning MOKE microscopy performed on different 1L, 2L, and 3L CrPS$_4$ flakes (Figure 3a). In these measurements, samples were subjected to a constant out-of-plane magnetic field and the beam spot relative to the sample position is scanned to obtain the spatial 2D map of the spin orientations. Note that uniformly positive $\theta_{KR}$ signals are detected only on 1L and 3L flakes. On the 2L flake, however, no net magnetization is found, showing the strong even-odd contrast. The edge signals found in the sample boundaries reflect the polarization disturbance of the probe beam when it passes across step edges.

The ferromagnetism in odd layer flakes is further confirmed by performing field-sweep MCD measurements on 1L and 3L $CrPS_4$. In the measurement, 1L and 3L samples are loaded in a magneto-optic cryostat kept at the temperature of 3.5 K with a sweeping field applied in the out-of-plane direction. As shown in Figure 3b, clear hysteresis curves are observed for 1L and 3L samples, which shows direct evidence of ferromagnetism in odd layer thin $CrPS_4$. For 3L sample, a relatively large coercive force is found which is about 90 mT with a near square shape of hysteresis loop, suggesting the material's high magneto-crystalline anisotropy. In 1L sample, the coercive force is found to be smaller than in 3L sample, showing the possibility of domain formation.

Moreover, $CrPS_4$ sample is found to be surprisingly stable under air exposure. Figure 4a is the transmission electron microscopy (TEM) data of bulk $CrPS_4$ measured right after exfoliation and after a day of air exposure, showing almost no change of material crystallinity as can be checked by the characteristic angle in agreement with the crystal structure (Figure 1a). We have also checked the magnetization signal of thin $CrPS_4$ before and after the air exposure as shown in Figure 4b. For this measurement, 1L, 3L, and 5L samples are first exfoliated in a glove box and moved to a cryostat while keeping the exposure to air minimal (less than 1 minute). Clear FM transition curves are found similar to the samples that are exfoliated in air (Figure 2). The same measurement is performed after venting the chamber and exposing the samples in air for about a day and similar transition curves and $T_C$ were reproduced, revealing the good air stability of net magnetization in thin CrPS$_4$. More detailed air stability test measurements are in Supporting Information Fig. S12.

We now proceed to the theoretical analysis of the ferromagnetism observed in CrPS$_4$ based on the first-principles density functional theory (DFT) calculations. From the DFT simulations, the origin of the ferromagnetism in monolayer CrPS$_4$ is found to be the occupied Cr $3d$ $t_{2g}$ ($d_{xy}$, $d_{yz}$, $d_{zx}$) states hybridized with S $3p$ in the upper part of the valence band. The calculation also supports alternating spin direction along the vertical axis for 2L and 3L systems since interlayer AFM coupling is energetically more stable than interlayer FM one (DFT calculations in Supporting Information). This has been verified from the spin projected density of states calculation (Figure 5a) and the negative value of the total energy difference, $E_{AFM} - E_{FM}$, which supports the AFM coupling between layers (Figure 5b). From our calculation, the interlayer coupling constant ($J_c$) is calculated to be a few hundreds of $\mu eV$ for 2L CrPS$_4$.

Lastly, we discuss the evolution of the inter-site interactions of CrPS$_4$ from bulk down to the monolayer. To consider both inter-site nearest-neighbor exchange couplings and out-of-plane anisotropy, we employ the spin Hamiltonian using the intralayer ($J_1$) and interlayer ($J_c$) coupling constants and magnetic anisotropy term ($K$):

$$H_{spin} = J_1 \sum_{i,j} \boldsymbol{S}_i \cdot \boldsymbol{S}_j + J_c \sum_{i,j} \boldsymbol{S}_i \cdot \boldsymbol{S}_j + K \sum_i (S_i^z)^2 - g\mu_B \boldsymbol{H} \cdot \sum_i \boldsymbol{S}_i, \qquad (1)$$

Here $\boldsymbol{S}$ is the spin operator, $\boldsymbol{H}$ is the external magnetic field, $i$ and $j$ are lattice sites, $g$ is the Lande $g$-factor, and $\mu_B$ is the Bohr magneton. To determine the parameters $J_1$, $J_c$, and $K$, we first experimentally obtained temperature and field dependent spin-flop fields and phases from a bulk crystal (Supporting Information Fig. S13). By fitting the experimental data with relations obtained from Eq. (1), the coupling constants and anisotropy of the bulk material are extracted: $J_1 =$

$-2.4\ meV$, $J_c = +0.15\ meV$, and $K = -0.0042\ meV$. More details can be found in the model Hamiltonian calculation in Supporting Information. Using these parameters, the magnetization is computed using the finite-temperature Monte-Carlo simulations from bulk down to the monolayer, which agrees well with experimental results (Supporting Information Fig. S15 and S16). The example of the computed transition curve for the monolayer is shown in the inset of Figure 5b. Thus, the exchange coupling constants of bulk is found to be consistently effective in thin materials, which enables the distinct even-odd contrasting layer-dependent magnetism in thin layers of CrPS$_4$ even after the exfoliation in ambient condition.

**CONCLUSIONS**

In conclusion, the van der Waals magnetic crystal with persistent spin configuration that is robust to exfoliation and air exposure will be useful for the controlled fabrication of heterostructures for studying stacking-dependent magnetic interface phenomena and Moiré superlattices.[27,28] Furthermore, the outstanding material stability found in van der Waals 2D ferromagnet will be useful for further investigation of the spin-dependent quantum and topological phenomena.[29-31] Our finding of the extreme air stable CrPS$_4$ would make various kinds of 2D magnetic heterostructures possible, providing an exciting venue for the van der Waals spintronics applications.

**METHODS**

**CrPS$_4$ crystal synthesis**

CrPS$_4$ single crystals were synthesized by chemical vapour transport method. The original compounds, chromium metallic powder (Alfa Aesar, >99.996%), phosphorous red (Sigma-Aldrich, >99.99%) and sulfur flakes (Sigma-Aldrich, >99.999%), were mixed in the Ar-filled glove box in the stoichiometric ratio with the additional 5% of sulfur for the agent and sealed inside a quartz tube. The sealed quartz tube was placed in the two-zone furnace to initiate the synthesis. The furnace was heated to 650/550 ℃ for 1 week and slowly cooled down to room temperature. The final material is crystallized in shiny and black plates. The stoichiometric ratio of the single crystal was further checked by an energy dispersive X-ray (EDX) spectroscopy.

**Thin sample preparation**

1L to 5L CrPS$_4$ flakes are obtained by exfoliating bulk synthesis crystals onto 285 nm thick SiO$_2$ layer on Si substrates. The thicknesses of the flakes are first identified by image contrasts in optical microscopy and confirmed by Raman spectroscopy and atomic force microscopy. All samples are exfoliated in ambient condition except samples used for air stability test. Samples used for air stability test are first exfoliated in a glove box to compare magnetization before and after the air exposure. Samples used for field-sweep experiments are prepared by covering the sample by h-BN layer to allow several days of time intervals between multiple number of measurements.

**Neutron powder diffraction**

Time-of-flight powder neutron diffraction was carried out at the SuperHRPD beamline of MLF, J-PARC.[32,33] The powder diffraction was performed at 150 and 3 K, above and below $T_N$.

SARAh[34] and Fullprof program[35] were used for magnetic symmetry analysis and Rietveld analysis, respectively.

**Raman spectroscopy**

A continuous-wave 532 nm laser with the beam power of ~ 170 µW was focused onto the sample by an objective (NA = 0.6). The focused beam diameter was ~ 1 µm. The reflected light is collected using the same objective lens, guided to the collection path by a beam splitter, and analyzed by a grating spectrometer (1,800 grooves/mm) equipped with a charge-coupled device (CCD). Two notch filters are used to remove the scattered pump light in the collection path. For the site-dependent measurement, the sample position was moved by an XY piezo-stage inside the cryostat.

**Magneto-optic Kerr effect microscopy**

In the magneto-optic Kerr effect (MOKE) measurement, a continuous-wave 532 nm laser is used as a probe beam with power no more than 100 µW. The polarization of the incidence probe light is adjusted by a linear polarizer and a half-wave plate (HWP) in the excitation path. The probe light is focused on the sample by an objective (NA = 0.6) under normal incidence. The reflected light collected using the same objective lens is passed through a photo-elastic modulator (PEM), which modulates the sign of the Kerr rotation (KR) angle at a frequency of 100 kHz. In the detection part, HWP and Wollaston prism were used to split the beam into two orthogonal linear polarizations. A pair of photodiodes are used to measure the balanced signal through lock-in detection with the reference frequency provided by PEM. 2D MOKE image is obtained by scanning an XY piezo-stage in the cryostat. To initialize the spin alignment in the temperature-dependent and 2D scanning measurement, a permanent magnet was placed beneath the sample inside a cryostat, exerting an out-of-plane field of ~ 0.25 T at the sample position.

**Magnetic circular dichroism**

Magnetic circular dichroism (MCD) measurements used 532 nm laser as a probe beam and samples were loaded either in a cryostat with a permanent magnet for fixed field measurements or in a magneto-optic cryostat equipped with a superconducting magnet for field-sweeping measurements. PEM is placed in the excitation path to create alternating right and left circularly polarized light with a frequency of 50 kHz, and lock-in detection is used to measure the absorbance difference between the two opposite circular polarizations of excitation source.

**Density functional theory calculation**

We carried out the first-principles pseudopotential total energy calculations based on the density functional theory (DFT) within the generalized gradient density approximation (GGA) as implemented in the Vienna *ab Initio* simulation package (VASP).[36-38] The electron-ion interactions were described employing the projector-augmented-wave (PAW) method[39] and Perdew-Burke-Ernzerhof (PBE) functional was used for GGA exchange correlation potential.[40] The GGA plus on-site Coulomb repulsion U (GGA+U) approach by Dudarev *et al.* was used to describe strongly localized Cr 3$d$ orbitals.[41] The effective on-site Coulomb interaction parameter ($U_{eff} = U - J = 2$ eV) was applied to Cr 3$d$ orbitals. The van der Waals dispersion correction is included using DFT-D3 method of Grimme as implemented in the VASP code.[42] For CrPS$_4$ slab models, we used a kinetic energy cutoff of 500 eV, and Monkhorst-pack k-mesh of $\Gamma$–centered 2x3x1 for Brillouin zone integration is employed.[43] All internal atomic positions and the lattice parameter were fully relaxed until the total energy was converged within $10^{-6}$ eV. A vacuum spacing of more than 18 Å was chosen to separate the periodic slabs in the out-of-plane direction.

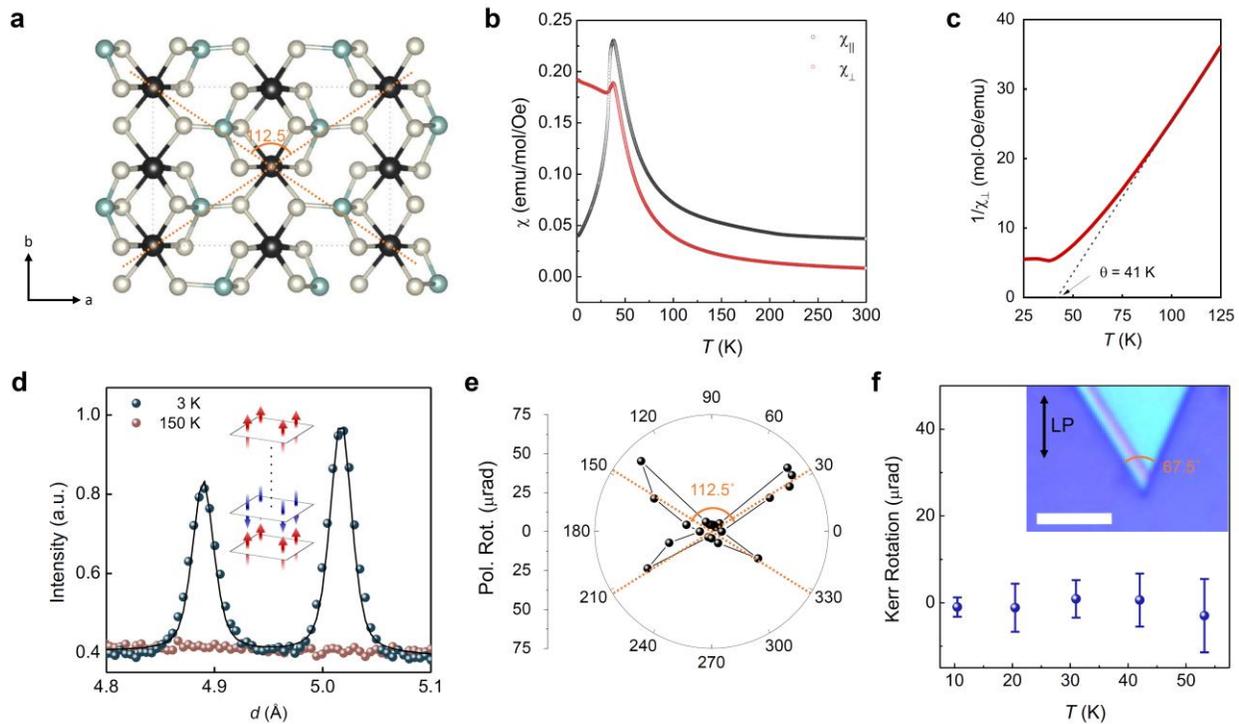

**Figure 1 | Crystal structure and magnetic properties of bulk CrPS₄. a.** Crystal structure of CrPS₄ in ab-plane. Black, cyan, and gray spheres are chromium, phosphorous, and sulfur atoms, respectively. **b.** Magnetic susceptibilities measured under the magnetic field of 100 Oe applied along ($\chi_\parallel$) and perpendicular ($\chi_\perp$) to the c-axis, respectively. **c.** Inverse of the susceptibility $\chi_\perp$. The Curie-Weiss temperature was fitted to +41 K from the $x$-intercept of the extrapolated data. **d.** Time-of-flight powder neutron diffraction data taken at 3 and 150 K. Simulated result at 3 K with the existence of (hk$\frac{1}{2}$) peaks is well-matched to the experimental data (black solid line), implying the A-type antiferromagnetic ordering of bulk CrPS₄ (inset). **d.** Polar plot of the optical rotation of CrPS₄ bulk crystal at room temperature. The polarization rotation is measured as a function of the relative angle between the incident probe polarization and a-axis of the crystal. Dotted guidelines and angles are shown for comparison with the crystal structure. **e.** MOKE signal as a function of temperature on bulk CrPS₄ under $\mu_0 H = 0.25$ T. Inset: Optical microscope image of bulk. LP refers to the orientation of the linear polarizer for the incident beam. The scale bar is 5 μm.

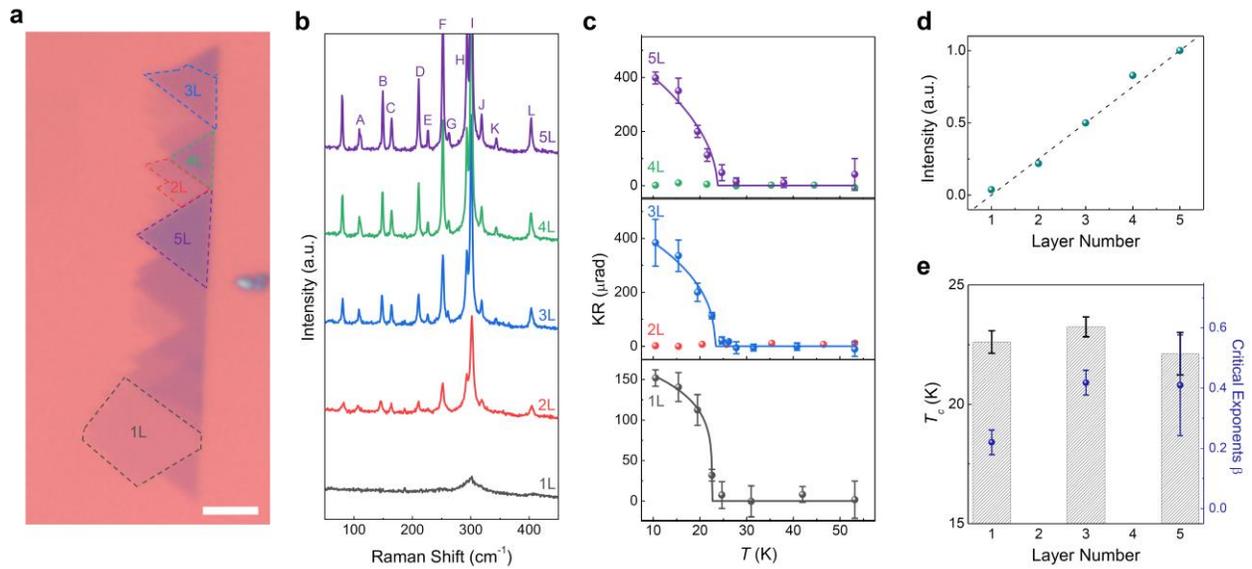

**Figure 2 | Layer dependent magnetic properties of 1L to 5L CrPS$_4$. a.** Microscope image of exfoliated CrPS$_4$. Layer numbers are indicated with colored dots. The scale bar is 5 μm. **b.** Raman spectra measured on 1L to 5L flakes shown in **a**. The peaks are labeled alphabetically from A to L in the order of increasing Raman shift. **c.** Temperature dependence of MOKE signal under $\mu_0H = 0.25$ T measured on 1L to 5L flakes shown in **a**. Top: 5L (purple) and 4L (green), Middle: 3L (blue) and 2L (red), Bottom: 1L (gray). Symbols and solid lines are experimental data and fit results using the function $\theta_{KR}(T) = \alpha \left(1 - \frac{T}{T_c}\right)^{\beta}$, respectively. **d.** Raman peak intensity of the F peak for each flake normalized by that from the 5L as a function of the layer number. A linear fit function (dashed line) is shown as a reference. **e.** Critical temperature, $T_C$ (black dashed bar), and critical exponent, $\beta$ (blue symbol), extracted from the fitting lines shown in **c**.

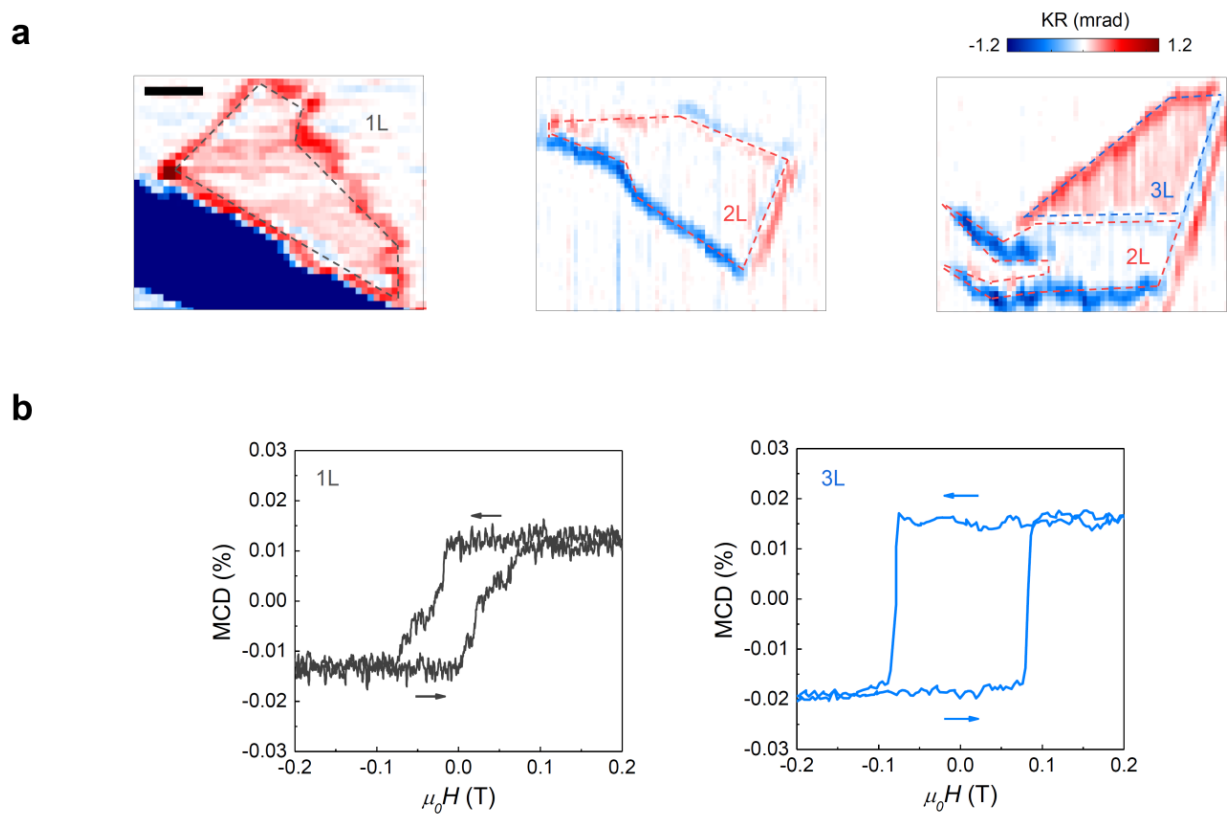

**Figure 3 | Ferromagnetism in 1L and 3L CrPS$_4$. a.** Scanning MOKE microscope images of 1L, 2L, and 3L CrPS$_4$ under $\mu_0 H = 0.25$ T at the temperature of 5 K. The scale bar is 2 μm. The flake boundaries are shown by dashed lines. **b.** Magnetic hysteresis loops of 1L (gray) and 3L (blue) samples measured by MCD at 3.5 K. Arrows show the field-sweep directions.

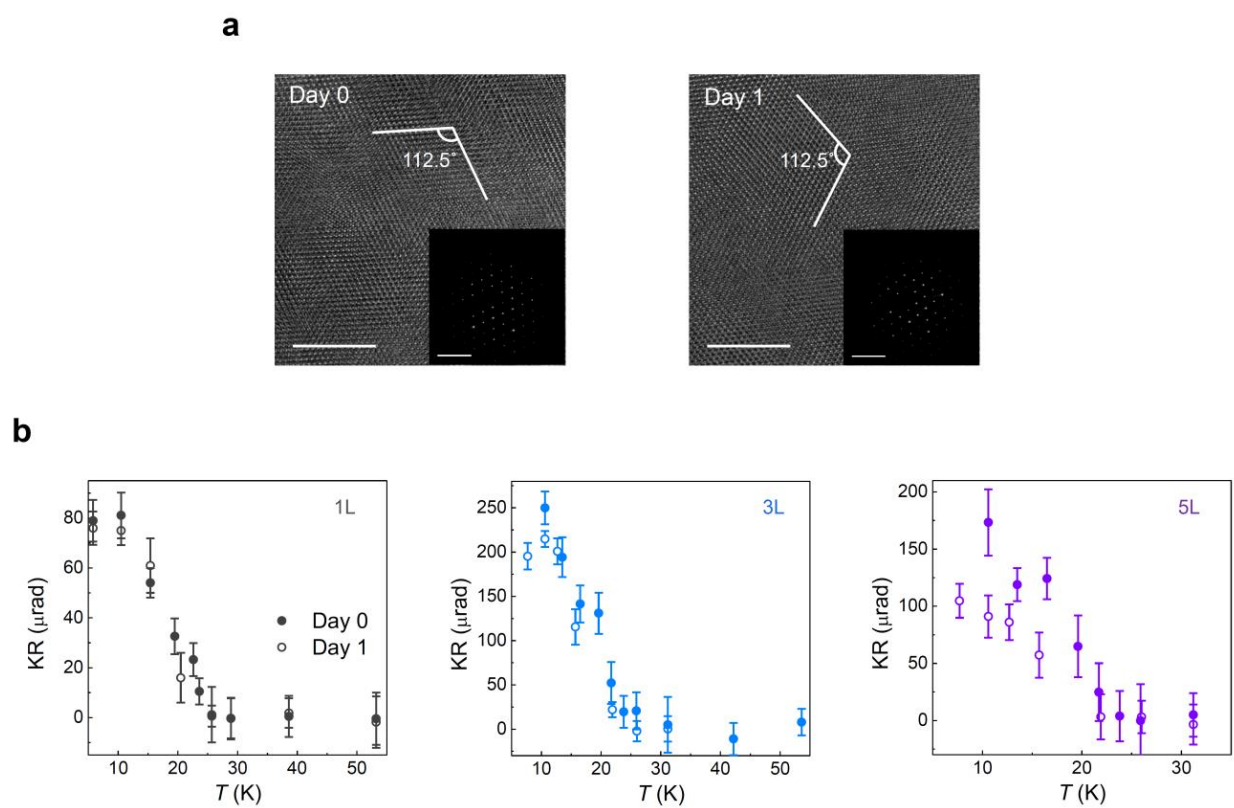

**Figure 4 | Air stability test measurements of CrPS₄. a.** TEM images of bulk CrPS₄ right after exfoliation and after 1 day of air exposure (Scale bars: 10 nm). Insets: Electron diffraction patterns (Scale bar: 5 1/nm). The characteristic angle of 112.5° indicating the good crystallinity of the sample is observed both before and after the air exposure. **b.** Temperature-dependent MOKE signal measured on 1L, 3L, and 5L samples with minimal air exposure (Day 0) and 1 day of air exposure (Day 1), showing stable magnetism of thin layer samples.

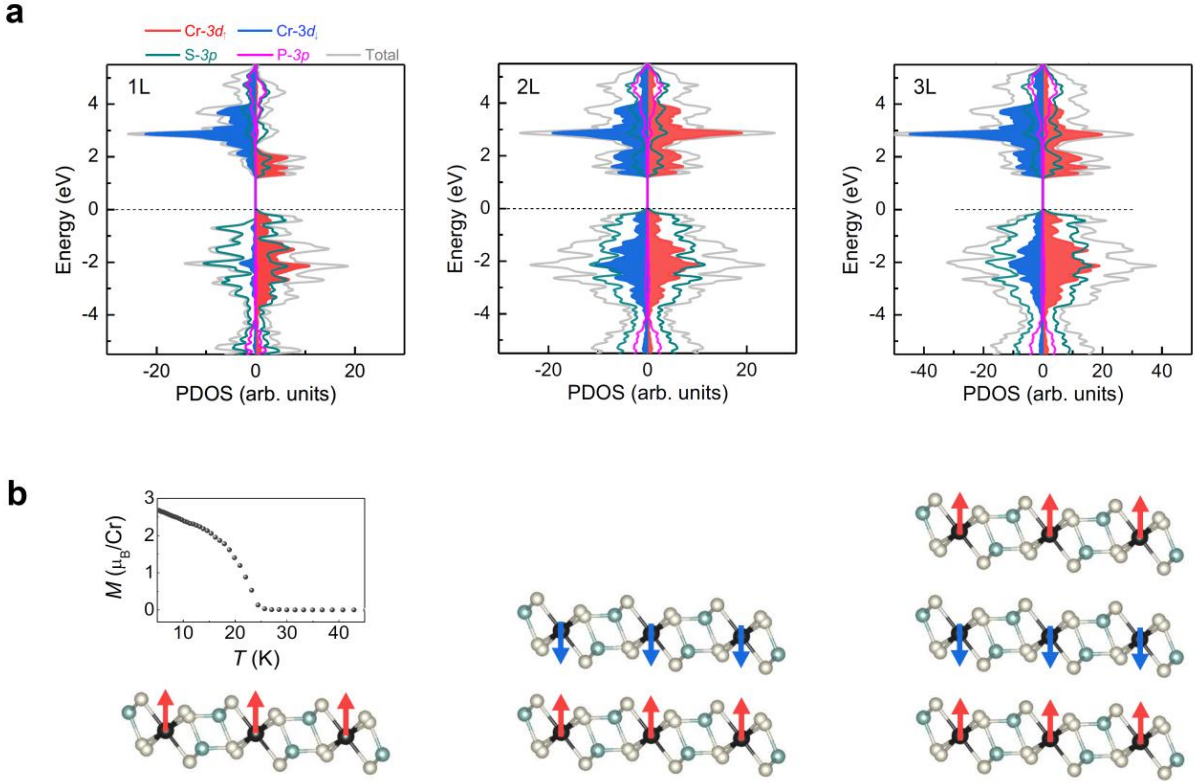

**Figure 5 | DFT calculations and spin alignments of 1L to 3L CrPS$_4$. a.** Orbital projected density of states (PDOS) calculated from 1L to 3L. The Cr 3$d$ DOS with spin up (red region) and spin down (blue region) states are separately drawn. The S 3$p$, P 3$p$, and total DOS are also shown by green, magenta, and gray solid lines, respectively. Valence band maximum is set to zero in PDOS. **b.** Magnetic orderings of 1L to 3L CrPS$_4$ with alternating spin up (red arrows) and spin down (blue arrows) layers. The total energy differences between interlayer AFM and FM state per formula unit, $\Delta E$ (= $E_{AFM} - E_{FM}$), obtained by DFT calculation for 2L and 3L are $\Delta E = -0.09$ meV/Cr and $\Delta E = -0.15$ meV/Cr, respectively. Inset: Monte-Carlo simulation of calculated magnetization in monolayer CrPS$_4$ as a function of temperature using the model Hamiltonian including the exchange interactions and anisotropy terms.

## ASSOCIATED CONTENT

### Supporting Information

The Supporting Information is available online: Neutron powder diffraction data, Atomic force microscopy of 1L to 5L $CrPS_4$, Raman spectroscopy of 1L to 5L $CrPS_4$, Angle dependent polarization rotation, Magnetic circular dichroism (MCD) measurement, Air stable magnetism in thin layer $CrPS_4$, Bulk SQUID-VSM measurement, DFT simulations of bulk and monolayer $CrPS_4$, Interlayer exchange coupling of 2L $CrPS_4$ using DFT results, Model Hamiltonian calculation; Monte-Carlo simulation of bulk and monolayer $CrPS_4$.

## AUTHOR INFORMATION

### Notes

The authors declare no competing financial interest.

## ACKNOWLEDGEMENTS

This work was supported by the New Faculty Startup Fund from Seoul National University. J.S., T.L. and J.L were supported from the National Research Foundation (NRF) of Korea (Grants No. 2017R1C1B2002631, No. 2020R1A2C2011334, No. 2020R1A5A6052558 and No. 2021R1A5A1032996). Work at IBS CCES was supported by Institute for Basic Science (IBS), Korea (Grant No. IBS-R009-G1). Work at CQM was supported by the Leading Researcher Program of the NRF, Korea (Grant No. 2020R1A3B2079375). Theoretical analysis of Maengsuk Kim and J.H.L. was supported by Creative Materials Discovery of the NRF, Korea (Grant No. 2017M3D1A1040828). J.O. and Miyoung Kim were supported by the National Research Foundation (NRF) of Korea (Grant No. 2017R1A2B3011629). Work at Cornell University was supported by the US National Science Foundation (NSF) under DMR-1807810. The SuperHRPD experiment was performed at the Materials and Life Science Experimental Facility of the J-PARC, Japan (Proposal No. 2019BF0802). J.L. acknowledges support from TJ Park Science Fellowship of POSCO TJ Park Foundation, Korea.